\documentclass[reqno]{amsart}

\usepackage[dvips]{graphicx}

\begin{document}

\title[Solitons of the Qiao equation]{Smooth soliton solutions of a new integrable equation by Qiao}

\author[S. Sakovich]{Sergei Sakovich}

\address{Institute of Physics, National Academy of Sciences, 220072 Minsk, Belarus \newline \indent Max Planck Institute for Mathematics, Vivatsgasse 7, 53111 Bonn, Germany}

\email{saks@tut.by}

\begin{abstract}
We find a transformation which relates a new third-order integrable nonlinear evolution equation, introduced recently by Qiao, with the well-known modified Korteweg--de~Vries equation. Then we use this transformation to derive smooth soliton solutions of the new equation from the known rational and soliton solutions of the old one.
\end{abstract}

\maketitle

\section{Introduction} \label{s1}

In the present paper, we study the following third-order integrable nonlinear evolution equation:
\begin{equation} \label{e1}
m_t = \left( \frac{1}{2 m^2} \right)_{xxx} - \left( \frac{1}{2 m^2} \right)_x .
\end{equation}
This equation was introduced recently by Qiao \cite{Q} as the second positive member in a new completely integrable hierarchy. The Qiao equation---we will refer to \eqref{e1} in this way---possesses a Lax representation and a bi-Hamiltonian structure \cite{Q,QL}. Moreover, the Qiao equation may be reduced from the two-dimensional Euler equation by an approximation procedure, and its solutions may be useful to construct new solvable potentials in Newtonian dynamics and to model electrophysiological phenomena in neuroscience \cite{Q}.

There is an interesting problem concerning the soliton solution of the Qiao equation. In \cite{Q,QL}, where the traveling wave solutions of \eqref{e1} were studied, it was stated that this equation has no smooth solitons, and only piecewise smooth, but not smooth, continuous traveling wave solutions were found in those papers. On the other hand, in \cite{LQ}, where the bifurcations of traveling wave solutions of \eqref{e1} were studied, it was stated that this equation does possess smooth soliton solutions, and some of those solitons were shown in Figure~5 of that paper. We believe that the complexity of investigation of the Qiao equation's solitons is caused mainly by the fact that they can be expressed only in a parametric form. In the present paper, we also address the problem on smooth soliton solutions of \eqref{e1}, but we do not use the traveling wave reduction to an ordinary differential equation. Instead, we follow the way used in \cite{SS} for the short pulse equation: we transform the studied new equation into a known old one, and then transform known explicit expressions for solutions of the old equation into sought parametric expressions for solutions of the new one.

The paper is organized as follows. In Section~\ref{s2}, we find a transformation which relates the Qiao equation with the well-known modified Korteweg--de~Vries (mKdV) equation. Then, in Sections~\ref{s3} and~\ref{s4}, we apply this transformation, respectively, to the rational solution and to the soliton solution of the mKdV equation, and in this way we obtain three types of smooth solitons of the Qiao equation as well as three types of its smooth antisolitons. Section~\ref{s5} contains some concluding remarks.

\section{Transformation to the mKdV equation} \label{s2}

Let us try to transform the new Qiao equation \eqref{e1} into one of well-studied old equations. We do this in the way used in \cite{S1} for the Fujimoto--Watanabe equations; see also \cite{S2,S3} for more theory and applications of the transformations involved.

First of all, we transform the nonlinear evolution equation \eqref{e1}, which has the form $m_t = - m^{-3} m_{xxx} + (\text{lower-order terms})$, into an equation of the form $u_t = u^3 u_{xxx} + (\text{lower-order terms})$. To do this, we change in \eqref{e1} the dependent variable $m$ as
\begin{equation} \label{e2}
m (x,t) = \frac{-1}{u (x,t)} ,
\end{equation}
the result being the polynomial evolution equation
\begin{equation} \label{e3}
u_t = u^3 u_{xxx} + 3 u^2 u_x u_{xx} - u^3 u_x .
\end{equation}
This equation is just a special case of the item (II.2) in the Fujimoto--Watanabe classification \cite{FW}, $u_t = u^3 u_{xxx} + 3 u^2 u_x u_{xx} + 4 \alpha u^3 u_x$ with the arbitrary constant $\alpha$ fixed to be $-1/4$. Therefore, according to \cite{S1}, it is possible to relate the Qiao equation, through its previously known form \eqref{e3}, with the Korteweg--de~Vries equation. We will see soon, however, that it is more convenient to transform the Qiao equation into the mKdV equation, in order to easily derive solutions of the former from solutions of the latter.

Next, we take into account that the evolution equation \eqref{e3} can be written as
\begin{equation} \label{e4}
u_t = u^2 \left( u u_{xx} + u_x^2 - \frac{1}{2} u^2 \right)_x ,
\end{equation}
which means, according to \cite{S1,S2}, that the transformation
\begin{equation} \label{e5}
x = v (y,t) , \qquad u (x,t) = v_y (y,t)
\end{equation}
relates \eqref{e3} with the third-order evolution equation
\begin{equation} \label{e6}
v_t = v_{yyy} - \frac{1}{2} v_y^3 .
\end{equation}
This point should be explained in more detail. It is easy to see that any function $v (y,t)$ satisfying \eqref{e6} determines via \eqref{e5} a function $u (x,t)$ satisfying \eqref{e3}. In the opposite direction, when we substitute the expressions \eqref{e5} to \eqref{e3}, we obtain a fourth-order equation for $v (y,t)$, which, owing to the form of \eqref{e4}, can be written as
\begin{equation} \label{e7}
\left( \frac{v_t - v_{yyy}}{v_y} + \frac{1}{2} v_y^2 \right)_y = 0
\end{equation}
and then can be integrated over $y$, the result being the third-order equation
\begin{equation} \label{e8}
v_t = v_{yyy} - \frac{1}{2} v_y^3 + f (t) v_y ,
\end{equation}
where $f (t)$ is an arbitrary function. This means that any solution $u (x,t)$ of \eqref{e3} is represented parametrically, via \eqref{e5} with $y$ being the parameter, by a solution $v (y,t)$ of \eqref{e8} with some function $f (t)$. However, we can always fix this function $f (t)$ to be zero, by the corresponding change of the parameter $y$, $y \mapsto y + \int f (t) d t$, which has no effect on the represented solution $u (x,t)$. Consequently, all solutions of \eqref{e3} are represented by all solutions of \eqref{e6} via the relations \eqref{e5}.

Finally, we introduce the new dependent variable $w$,
\begin{equation} \label{e9}
w (y,t) = v_y (y,t) ,
\end{equation}
which is convenient because the relation
\begin{equation} \label{e10}
u (x,t) = w (y,t)
\end{equation}
holds due to \eqref{e5}. According to \eqref{e6} and \eqref{e9}, the function $w (y,t)$ is any solution of the mKdV equation
\begin{equation} \label{e11}
w_t = w_{yyy} - \frac{3}{2} w^2 w_y ,
\end{equation}
whereas the function $v (y,t)$ is determined by $w (y,t)$ via the relations $v_y = w$ and $v_t = w_{yy} - \frac{1}{2} w^3$ up to an arbitrary additive constant of integration. Of course, there is no need to further transform \eqref{e11} into the Korteweg--de~Vries equation, because it would only complicate the already found way to solve the Qiao equation.

We can summarize the obtained transformation, which relates the equations \eqref{e1} and \eqref{e11} to each other, as follows. All solutions of the Qiao equation \eqref{e1} are represented parametrically by all solutions of the mKdV equation \eqref{e11} via the relations
\begin{align}
m (x,t) &= \frac{-1}{w (y,t)} , \label{e12} \\
x &= v (y,t) : \quad \left\{
\begin{aligned}
v_y &= w , \\
v_t &= w_{yy} - \frac{1}{2} w^3 ,
\end{aligned}
\right. \label{e13}
\end{align}
where the independent variable $y$ serves as the parameter. Of course, this transformation is local, in the sense that it maps a local solution of \eqref{e11} to a local solution of \eqref{e1}, and for every particular $w (y,t)$ some extra analysis is required to see whether the corresponding $m (x,t)$ is real-valued, smooth, and global.

In the next two sections, in order to show how the transformation \eqref{e12}--\eqref{e13} works, we apply it to two simplest nontrivial solutions of the mKdV equation \eqref{e11}: the rational solution
\begin{equation} \label{e14}
w (y,t) = a + \frac{4 a}{a^2 r^2 - 1} , \qquad r = y - \frac{3}{2} a^2 t ,
\end{equation}
where $a$ is an arbitrary nonzero constant, and the soliton solution
\begin{equation} \label{e15}
w (y,t) = a + \frac{2 k^2}{\pm \sqrt{a^2 - k^2} \cosh (k s) - a} , \qquad s = y - \left( \frac{3}{2} a^2 - k^2 \right) t ,
\end{equation}
where $a$ and $k$ are arbitrary constants, $k \neq 0$ and $a^2 - k^2 \neq 0$. One can notice that the expressions \eqref{e14} and \eqref{e15} are somewhat different from the ones well known in the literature \cite{AS}. The reason is that the mKdV equation is usually studied in its focusing form, with a positive coefficient at the nonlinear term, whereas we have the defocusing case in \eqref{e11}, and the expressions \eqref{e14} and \eqref{e15} can be obtained from the well-known ones through a complex-valued scale transformation of the dependent variable and a change of notations for the constants.

\section{Transformation of the rational solution} \label{s3}

Let us derive a solution of the Qiao equation \eqref{e1} from the rational solution \eqref{e14} of the mKdV equation \eqref{e11} by means of the transformation \eqref{e12}--\eqref{e13}. We substitute the function $w (y,t)$ given by \eqref{e14} to the relations \eqref{e12} and \eqref{e13}, make the integration required in \eqref{e13}, and fix the arbitrary additive constant of integration so that to have $v = 0$ for $y = t = 0$. In this way, we obtain the following expressions:
\begin{align}
m (x,t) &= \frac{1 - a^2 r^2}{3 a + a^3 r^2} , \label{e16} \\
x &= a^3 t + a r - 4 \tanh^{-1} (a r) , \label{e17}
\end{align}
where $\tanh^{-1}$ stands for the hyperbolic arctangent. These expressions contain $y$ via the variable $r = y - \frac{3}{2} a^2 t$ only, therefore we can use $r$ instead of $y$ as the parameter, for simplicity. The function $m (x,t)$, determined parametrically by \eqref{e16}--\eqref{e17} with $r$ being the parameter, does satisfy \eqref{e1}, as one can prove by direct computations.

Until now, we worked with the solution \eqref{e14} in a purely algebraic way, and the quantities $a$, $y$, $t$ and $w$ could be even complex-valued there. However, in order that the obtained solution $m (x,t)$ \eqref{e16}--\eqref{e17} be a real-valued function of real-valued variables, the quantities $a$, $y$ and $t$ must be real-valued, and the condition
\begin{equation} \label{e18}
-1 < a r < 1
\end{equation}
must be satisfied, which is caused by the presence of $\tanh^{-1}$ in \eqref{e17}. The domain of variation of $x$ is $- \infty < x < \infty$ due to \eqref{e17} and \eqref{e18}. The function $m (x,t)$, due to \eqref{e16} and \eqref{e18}, is everywhere smooth; it is positive, with one maximum equal to $1/(3a)$, when $a > 0$; it is negative, with one minimum equal to $1/(3a)$, when $a < 0$; and its asymptotes at $x \to \pm \infty$ are zero.

\begin{figure}
\centering
\includegraphics[width=0.72\textwidth]{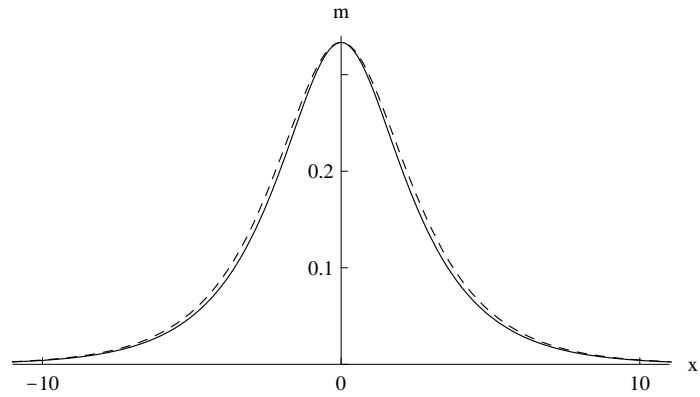}
\caption{The soliton solution \eqref{e16}--\eqref{e18} with $a = 1$ and $t = 0$ (solid), compared to the curve $m = 1 / \left( 3 \cosh \frac{x}{2} \right)$ (dashed).}
\label{f1}
\end{figure}

\begin{figure}
\centering
\includegraphics[width=0.72\textwidth]{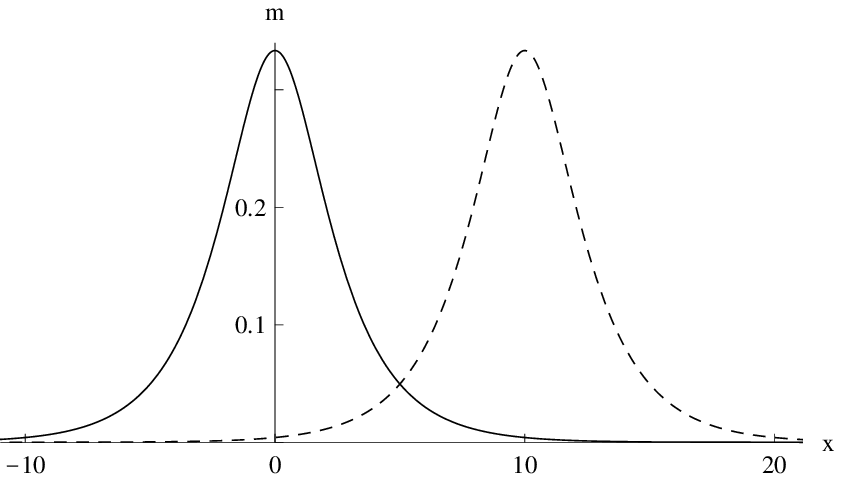}
\caption{The moving soliton \eqref{e16}--\eqref{e18} with $a = 1$: $t = 0$ (solid) and  $t = 10$ (dashed).}
\label{f2}
\end{figure}

\begin{figure}
\centering
\includegraphics[width=0.72\textwidth]{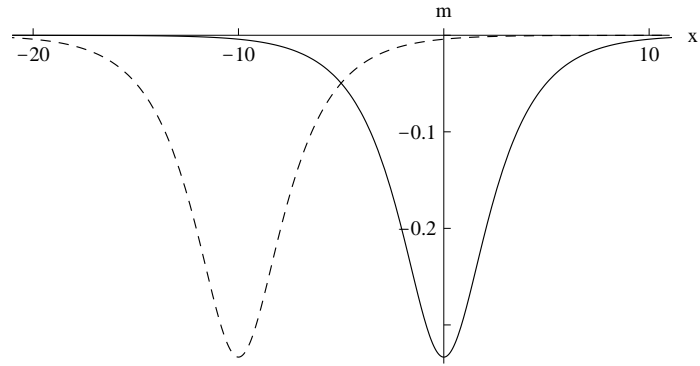}
\caption{The moving antisoliton \eqref{e16}--\eqref{e18} with $a = -1$: $t = 0$ (solid) and  $t = 10$ (dashed).}
\label{f3}
\end{figure}

We can see easily that the obtained solution \eqref{e16}--\eqref{e18} is a traveling wave solution, $m (x,t) = F \left( x - a^3 t \right)$ with the function $F$ being determined parametrically. The shape of this solitary wave is close to the shape of a hyperbolic secant curve but not identical to it, as shown in Figure~\ref{f1}. The speed of this solitary wave is equal to $a^3$; solitons (positive waves with $a > 0$) move to the right, and antisolitons (negative waves with $a < 0$) move to the left, as shown in Figures~\ref{f2} and~\ref{f3}, respectively. This antisymmetry $m \mapsto - m , \, x \mapsto - x$ between positive and negative solutions of the Qiao equation \eqref{e1} corresponds, via the transformation \eqref{e12}--\eqref{e13}, to the symmetry $w \mapsto - w$ of the mKdV equation \eqref{e11}.

To conclude the section, we point out the following two discovered phenomena. First, a rational solution of one nonlinear equation is transformed into a smooth soliton solution of another nonlinear equation. Second, only a part of the rational solution \eqref{e14} of the mKdV equation---the part bounded by the condition \eqref{e18}---is used to produce the entire smooth real solution of the Qiao equation.

\section{Transformation of the soliton solution} \label{s4}

Now, let us proceed to transformation of the soliton solution \eqref{e15} of the mKdV equation \eqref{e11}. The cases of ``$+$'' and ``$-$'' signs in \eqref{e15} will be considered separately.

We substitute the function $w (y,t)$ \eqref{e15}, with the choice of ``$+$'' sign in it, to the relations \eqref{e12} and \eqref{e13}, make the integration required in \eqref{e13}, fix the constant of integration so that to have $v = 0$ for $y = t = 0$, and obtain in this way the following expressions:
\begin{align}
m (x,t) &= \frac{a - \sqrt{a^2 - k^2} \cosh (k s)}{2 k^2 - a^2 + a \sqrt{a^2 - k^2} \cosh (k s)} , \label{e19} \\
x &= a \left( a^2 - k^2 \right) t + a s - 4 \tanh^{-1} \left( \frac{1}{k} \left( a + \sqrt{a^2 - k^2} \right) \tanh \frac{k s}{2} \right) . \label{e20}
\end{align}
These expressions contain $y$ via the variable $s = y - \left( \frac{3}{2} a^2 - k^2 \right) t$ only. Therefore we can use $s$ instead of $y$ as the parameter, for simplicity, when we consider \eqref{e19}--\eqref{e20} as a parametrically determined solution $m (x,t)$ of the Qiao equation \eqref{e1}. This is a traveling wave solution, $m (x,t) = G \left( x - a \left( a^2 - k^2 \right) t \right)$ with the function $G$ being determined parametrically, and the speed of the wave is equal to $a \left( a^2 - k^2 \right)$. Of course, in order that the obtained solution be a real solution, the quantities $a$, $k$, $s$ and $t$ must be real-valued in the right-hand sides of \eqref{e19} and \eqref{e20} (we take $k > 0$ without loss of generality), and the condition $k < |a|$ must be satisfied due to the presence of the square root. Further constraints on $s$ and $k$, caused by the presence of $\tanh^{-1}$ in \eqref{e20} and the denominator in \eqref{e19}, depend on the sign of $a$.

\begin{figure}
\centering
\includegraphics[width=0.72\textwidth]{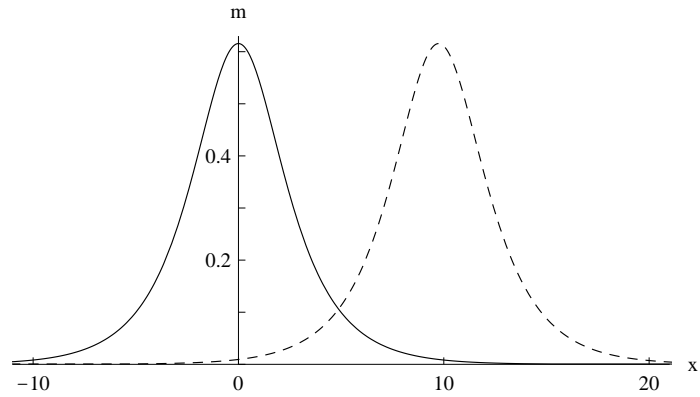}
\caption{The moving soliton \eqref{e19}--\eqref{e21} with $a = 1$ and $k = 0.95$: $t = 0$ (solid) and $t = 100$ (dashed).}
\label{f4}
\end{figure}

\begin{figure}
\centering
\includegraphics[width=0.72\textwidth]{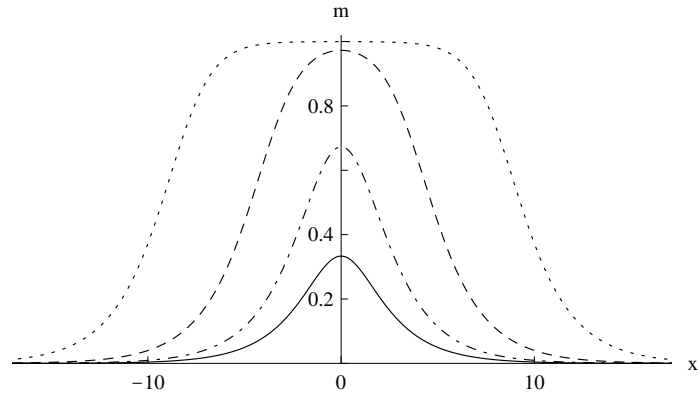}
\caption{The shape of the soliton \eqref{e19}--\eqref{e21} with $a = 1$ and $t = 0$: $k = 0.0001$ (solid), $k = 0.97$ (dot-dashed), $k = 0.9999$ (dashed), and $k = 1 - 10^{-8}$ (dotted).}
\label{f5}
\end{figure}

\begin{figure}
\centering
\includegraphics[width=0.72\textwidth]{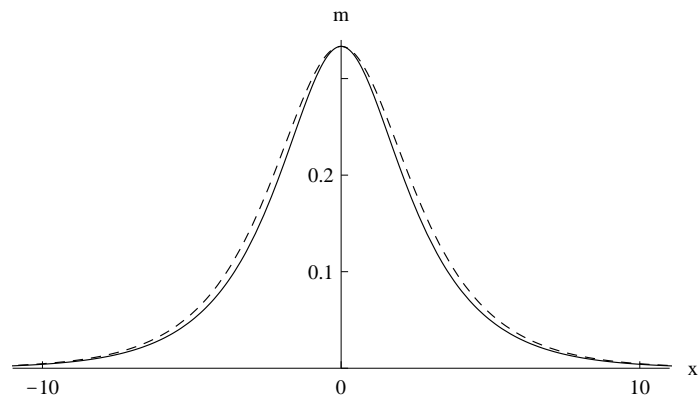}
\caption{The shape of the soliton \eqref{e19}--\eqref{e21} at $t = 0$: $a = 1$ and $k = 0.0001 a$ (solid); $a = 2.0186$ and $k = 0.97 a$ (dashed).}
\label{f6}
\end{figure}

When $a > 0$, the parameter $s$ must lie in the interval
\begin{equation} \label{e21}
|s| < \frac{2}{k} \tanh^{-1} \frac{k}{a + \sqrt{a^2 - k^2}} , \qquad 0 < k < a ,
\end{equation}
in order that $x$ \eqref{e20} be real-valued. With this constraint on $s$, the domain of variation of $x$ is $- \infty < x < \infty$, and the function $m (x,t)$ \eqref{e19} is everywhere smooth and positive, with one maximum equal to $1 / \left( a + 2 \sqrt{a^2 - k^2} \right)$ and zero asymptotes at $x \to \pm \infty$. The soliton, determined by \eqref{e19}--\eqref{e21}, moves with the speed $a \left( a^2 - k^2 \right)$ to the right, as shown in Figure~\ref{f4}. When $k \to 0$, the height and the speed of this soliton tend to $1/(3a)$ and $a^3$, respectively. In fact, the soliton \eqref{e16}--\eqref{e18} with $a > 0$, derived from the rational solution of the mKdV equation in Section~\ref{s3}, is a limiting case of the soliton \eqref{e19}--\eqref{e21} with $k \to 0$, as one can prove easily. On the other hand, when $k \to a$, the height and the speed of the soliton \eqref{e19}--\eqref{e21} tend to $1/a$ and $0$, respectively. Figure~\ref{f5} shows how the shape of the obtained soliton depends on the ratio $k/a$. We see that the soliton's top widens and flattens, when $k$ is very close to $a$, that is when the soliton's motion is extremely slow. However, the soliton's shape changes very little, when $k$ runs over the wide interval $0 < k \leq 0.97 a$. This is shown in Figure~\ref{f6}, where we have adjusted the values of $a$ so that to have solitary waves of equal heights. It is interesting that, of the two solitary waves shown there, the one which is only a little bit wider moves more than two times slower than the other one.

\begin{figure}
\centering
\includegraphics[width=0.72\textwidth]{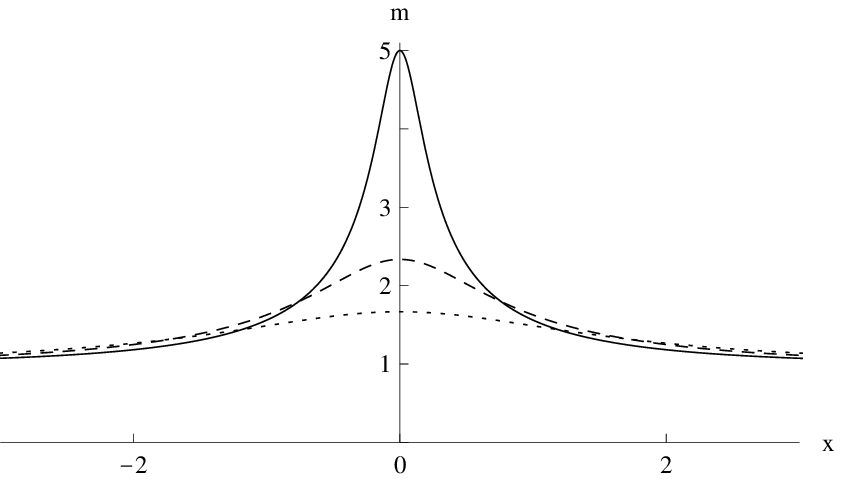}
\caption{The shape of the soliton, determined by \eqref{e19}, \eqref{e20} and \eqref{e22}, with $a = -1$ and $t = 0$: $k = 0.8$ (solid), $k = 0.7$ (dashed), and $k = 0.6$ (dotted).}
\label{f7}
\end{figure}

\begin{figure}
\centering
\includegraphics[width=0.72\textwidth]{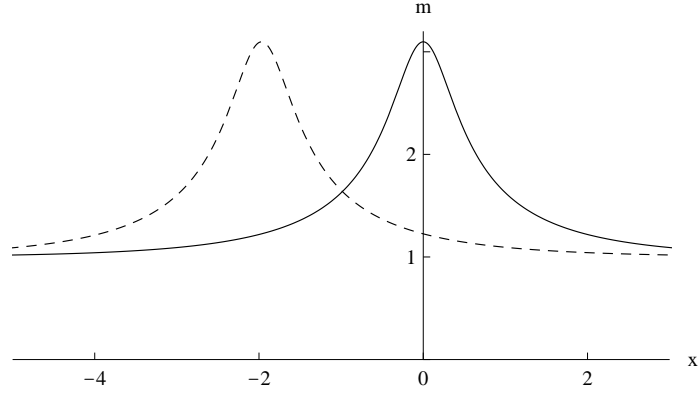}
\caption{The moving soliton, determined by \eqref{e19}, \eqref{e20} and \eqref{e22}, with $a = -1$ and $k = 0.75$: $t = 0$ (solid) and $t = 4.5$ (dashed).}
\label{f8}
\end{figure}

When $a < 0$ in \eqref{e19} and \eqref{e20}, the variable $x$ \eqref{e20} is real-valued on the whole interval $- \infty < s < \infty$ for any $k$, $0 < k < - a$, and the domain of variation of $x$ is $- \infty < x < \infty$, but the function $m (x,t)$ \eqref{e19} has one or two singularities if $k = - \sqrt{3} a / 2$ or $k > - \sqrt{3} a / 2$, respectively. The expressions \eqref{e19} and \eqref{e20} with
\begin{equation} \label{e22}
0 < k < - \frac{\sqrt{3}}{2} a , \qquad - \infty < s < \infty
\end{equation}
do determine a smooth soliton. This solitary wave is positive, has one maximum equal to $1 / \left( a + 2 \sqrt{a^2 - k^2} \right)$ and nonzero asymptotes $-1/a$ at $x \to \pm \infty$, and moves with the speed $a \left( a^2 - k^2 \right)$, that is to the left, as shown in Figures~\ref{f7} and~\ref{f8}. When $k \to 0$, the height of the soliton tends to $-1/a$, which is the level of the asymptotes, and the speed of the soliton tends to $a^3$. When $k \to - \sqrt{3} a / 2$, the soliton's height and speed tend to $+ \infty$ and $a^3/4$, respectively.

\begin{figure}
\centering
\includegraphics[width=0.72\textwidth]{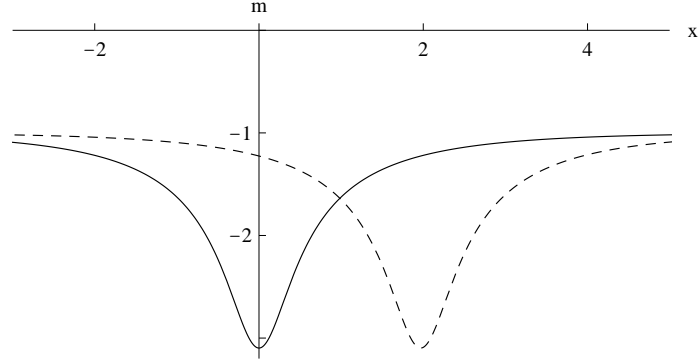}
\caption{The moving antisoliton \eqref{e23}--\eqref{e25} with $a = 1$ and $k = 0.75$: $t = 0$ (solid) and $t = 4.5$ (dashed).}
\label{f9}
\end{figure}

\begin{figure}
\centering
\includegraphics[width=0.72\textwidth]{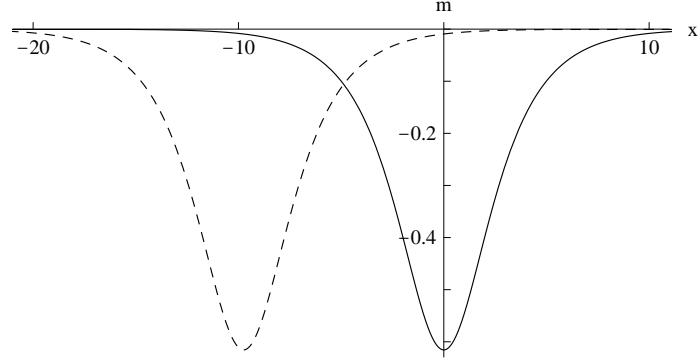}
\caption{The moving antisoliton, determined by \eqref{e23}, \eqref{e24} and \eqref{e26}, with $a = -1$ and $k = 0.95$: $t = 0$ (solid) and $t = 100$ (dashed).}
\label{f10}
\end{figure}

The case of ``$-$'' sign in the mKdV equation's solution \eqref{e15} is processed in the same way as the case of ``$+$'' sign. As the result, we obtain the following expressions:
\begin{align}
m (x,t) &= \frac{a + \sqrt{a^2 - k^2} \cosh (k s)}{2 k^2 - a^2 - a \sqrt{a^2 - k^2} \cosh (k s)} , \label{e23} \\
x &= a \left( a^2 - k^2 \right) t + a s - 4 \tanh^{-1} \left( \frac{1}{k} \left( a - \sqrt{a^2 - k^2} \right) \tanh \frac{k s}{2} \right) . \label{e24}
\end{align}
In order to get real, smooth and global solutions of the Qiao equation \eqref{e1}, we must supplement these expressions either with
\begin{equation} \label{e25}
0 < k < \frac{\sqrt{3}}{2} a , \qquad - \infty < s < \infty
\end{equation}
for positive values of $a$, or with
\begin{equation} \label{e26}
|s| < \frac{2}{k} \tanh^{-1} \frac{k}{- a + \sqrt{a^2 - k^2}} , \qquad 0 < k < - a
\end{equation}
for negative values of $a$. The obtained solitary waves are negative. The antisoliton \eqref{e23}--\eqref{e25} moves to the right and has nonzero asymptotes, as shown in Figure~\ref{f9}. It corresponds to the soliton, determined by \eqref{e19}, \eqref{e20} and \eqref{e22}, in the sense of the antisymmetry between positive and negative solutions, discussed in Section~\ref{s3}. The antisoliton, determined by \eqref{e23}, \eqref{e24} and \eqref{e26}, moves to the left and has zero asymptotes, as shown in Figure~\ref{f10}. It corresponds to the soliton \eqref{e19}--\eqref{e21} in the sense of that antisymmetry. Owing to this correspondence, there is no need to discuss these antisolitons in more detail.

To conclude the section, we remind the following important observation. When we used the soliton solution \eqref{e15} of the mKdV equation to obtain smooth real solutions of the Qiao equation, we had to impose certain constraints on the quantities $y$, $t$, $a$ and $k$ in the original solution, otherwise the obtained solutions would be singular or complex.

\section{Conclusion} \label{s5}

In this paper, we found a transformation which relates a new integrable nonlinear evolution equation of Qiao with the well-known mKdV equation. By means of that transformation, we derived smooth soliton solutions of the Qiao equation from the known rational and soliton solutions of the mKdV equation. We obtained three types of positive solitary wave solutions, or solitons. Solitons of two types have zero asymptotes and move to the right, whereas solitons of the third type have nonzero asymptotes and move to the left. Also we obtained three types of negative solitary wave solutions, or antisolitons, which have the same shape as the corresponding solitons but differ from them in their sign and direction of motion.

It is remarkable that the rational solution of the mKdV equation is transformed into a smooth soliton solution of the Qiao equation. An important detail of this is that only a part of the rational solution of the mKdV equation produces the entire smooth soliton solution of the Qiao equation. When the soliton solution of the mKdV equation is transformed, it is also necessary to impose certain constraints on its variables and constants, in order that the obtained solution of the Qiao equation be real and smooth. These observations may be important for further development of the method to obtain parametric representations for solutions of nonlinear wave equations via transformations of the equations.

Last but not least, we believe that the discovered transformation will be useful to obtain and study multisoliton solutions of the Qiao equation with the help of the Hirota bilinear representation of the mKdV equation \cite{H}.

\section*{Acknowledgements}

This work was partially supported by the BRFFR grant $\Phi$10-117. The author also thanks the Max Planck Institute for Mathematics for hospitality and support.

\end{document}